\documentclass[reprint,amsmath,amssymb,aps,pra,longbibliography,citeautoscript,secnumarabic,superscriptaddress]{revtex4-1}

\usepackage{graphicx}
\usepackage{dcolumn}
\usepackage{bm}
\usepackage[english]{babel}
\usepackage{color}
\usepackage{braket}
\usepackage{latexsym}
\usepackage{bbold}
\usepackage{epstopdf}
\usepackage[utf8]{inputenc}
\newcommand{\dmat}{\mathfrak r}

\addto\captionsenglish{}

\begin{document}

\title{Gravitationally induced phase shift on a single photon}

\author{Christopher Hilweg}
\email{christopher.hilweg@univie.ac.at}
\affiliation{
Faculty of Physics, University of Vienna, Boltzmanngasse 5, A-1090 Vienna, Austria
}
\author{Francesco Massa}
\affiliation{
Faculty of Physics, University of Vienna, Boltzmanngasse 5, A-1090 Vienna, Austria
}
\author{Denis Martynov}
\affiliation{LIGO Laboratory, Massachusetts Institute of Technology, Cambridge, MA 02139, USA}
\author{Nergis Mavalvala}
\affiliation{LIGO Laboratory, Massachusetts Institute of Technology, Cambridge, MA 02139, USA}
\author{Piotr T. Chru$\acute{s}$ciel}
\affiliation{Erwin Schrödinger Institute and Faculty of Physics, University of Vienna, A-1090 Vienna, Austria}
\author{Philip Walther}
\affiliation{
Faculty of Physics, University of Vienna, Boltzmanngasse 5, A-1090 Vienna, Austria
}

\date{\today}
\begin{abstract}
The effect of the Earth's gravitational potential on a quantum wave function has only been observed for massive particles. In this paper we present a scheme to measure a gravitationally induced phase shift on a single photon travelling in a coherent superposition along different paths of an optical fiber interferometer. To create a measurable signal for the interaction between the static gravitational potential and the wave function of the photon, we propose a variant of a conventional Mach-Zehnder interferometer. We show that the predicted relative phase difference of $10^{-5}$ radians is measurable even in the presence of fiber noise, provided additional stabilization techniques are implemented for each arm of a large-scale fiber interferometer. Effects arising from the rotation of the Earth and the material properties of the fibers are analysed. We conclude that optical fiber interferometry is a feasible way to measure the gravitationally induced phase shift on a single-photon wave function, and thus provides a means to corroborate the equivalence of the energy of the photon and its effective gravitational mass. 
\end{abstract}
\maketitle

\section{INTRODUCTION}
\label{Sec:Introduction}

Interferometry has proven to be an effective tool for high-sensitivity measurements in physics. For example, the recent groundbreaking detection of gravitational waves \cite{Abbott} relied on Michelson interferometers to measure ripples in the curvature of space-time predicted 100 years ago by the theory of General Relativity (GR). This achievement indicates that state-of-the-art technology enables the realization of interferometers capable of detecting gravitational effects even on quantum particles. While there exists a well developed framework incorporating general relativistic gravity into quantum physics \textemdash quantum field theory on curved space-time \cite{Birrell82}\textemdash its most distinctive predictions, such as Hawking radiation from black holes, are nowhere close to be testable in the near future. Various experiments have been proposed to test its other predictions, e.g. general relativistic corrections to the Newtonian gravitational phase shift for massive particles \cite{Wajima97,Dimopoulos08}. Yet, the size of the interferometric setups that are necessary to observe these corrections is still beyond the reach of the present-day technological capabilities. Rapidly advancing quantum optics technology allows for quantum states of light to be transmitted over increasingly large distances,
which sparked proposals for experiments probing the effects of the space-time curvature on photons \cite{Rideout12}. As a first important step towards this goal, we propose an experimental scheme for observing the gravitational phase shift on a single photon. We discuss the gravitational effects on the single-photon state inside a Mach-Zehnder interferometer (MZI), which is placed vertically inside the Earth's gravitational field. We show that the use of optical fibers together with a modification of the classic Mach-Zehnder scheme allows to obtain a detectable signal, even in the presence of noise.\\

The first experiment testing the influence of a gravitational potential at a quantum level was performed by Colella, Overhauser and Werner using neutrons in a matter-wave interferometer \cite{Werner75}. Subsequent measurements of the gravitational acceleration \textit{g} have also relied on atoms and molecules \cite{Cronin09}. However, because these experiments all used massive particles, they can be interpreted within the framework of Newtonian gravity; on the other hand, gravitational tests with massless particles, e.g. measurement of the Shapiro delay \cite{Shapiro64}, require a general relativistic explanation \cite{Zych11}.
In order to detect GR effects on single photons, a MZI with arms located at different heights above the Earth can be used \cite{Zych11,Zych12}. The time of emission of the photon serves as a \textquoteleft clock\textquoteright, keeping track of the evolved proper time along each path of the interferometer \cite{Zych11}. By reading the state of the clock we gain information about the path taken by the photon inside the MZI, which should lead to a drop in visibility according to the quantum complementarity principle \cite{Englert96,Bohr1928}. We therefore expect the visibility to start decreasing once the relative GR proper time difference between the arms approaches the photon coherence time (the precision of the clock). The observation of such a drop in visibility would constitute a genuine test of the interplay between GR and quantum mechanics. Unfortunately, interferometers with arm lengths of a few thousand kilometers would be required for such an experiment. Such large-scale interferometers are currently only feasible for space-based experiments \cite{Pallister16}. However, a smaller interferometer still allows for measuring gravitationally induced phase shifts as we discuss in this work. A successful demonstration of this phase shift would constitute e.g. a verification of the equivalence between the energy of a single photon and its effective gravitational mass \cite{Zych12}.\\

\begin{figure}
    \includegraphics[scale=0.35]{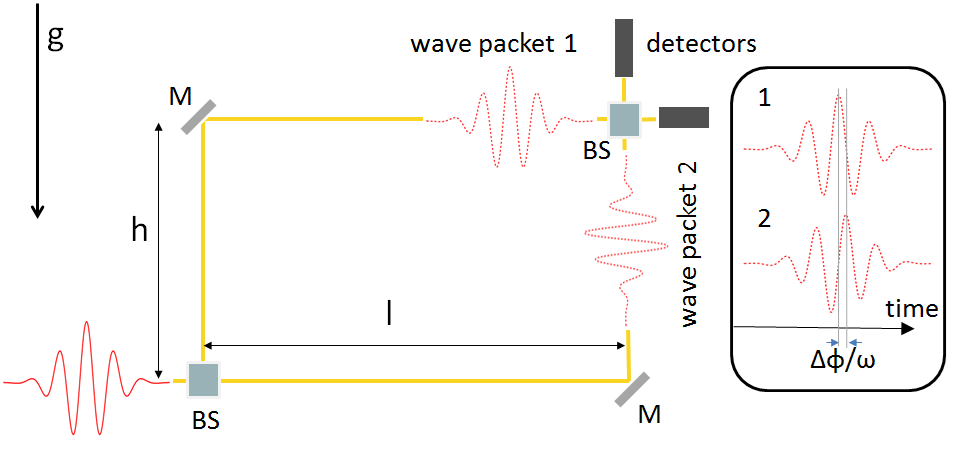}
    \caption{Schematic of a conventional Mach-Zehnder interferometer of area $A=lh$ placed vertically inside Earth's gravitational field. A beam splitter (BS) transforms a single photon into a coherent superposition between the two possible arms. The wave function evolves along two different paths and arrives at the merging beam splitter at the same time but slightly shifted in phase due to the presence of a gravitational potential.}
    \label{OrdinaryMZI}
\end{figure}

A conventional MZI, with arms of length \textit{l} separated in height by a distance \textit{h}, placed vertically inside Earth's gravitational field is illustrated in Fig. \ref{OrdinaryMZI}. The relation between the enclosed area $A=lh$ and gravitational phase difference is approximately given by \cite{Zych12}

\begin{equation}
\label{PhaseShift}
\Delta \phi_g \approx \frac{2\pi A N g}{\lambda c^2}\,,
\end{equation}
where \textit{N} is the group effective index of the transmissive medium for light propagation, \textit{g} is the gravitational acceleration, $\lambda$ is the central wavelength used to excite the interferometer and \textit{c} is the speed of light in vacuum. The gravitational phase shift can be interpreted to be a result of the coupling of the average energy of a photon with effective mass $m=\frac{h\nu_0}{c^2}=\frac{h}{c\lambda}$ to the Newtonian gravitational potential \cite{Zych12}. If the coherence time of the photon is much larger than the proper time difference experienced between different paths the fringe visibility is high and the detection probabilities at the two output detectors are given by

\begin{equation}
\label{DetProb}
P_{\pm} = \frac{1}{2} \Big( 1 \pm \cos(\Delta \phi_g+\phi(t)) \Big),
\end{equation}
where $\phi(t)$ denotes the time varying phase-noise contributions present in any real interferometric setup.

\begin{figure*}
\centering
    \includegraphics[scale=0.6]{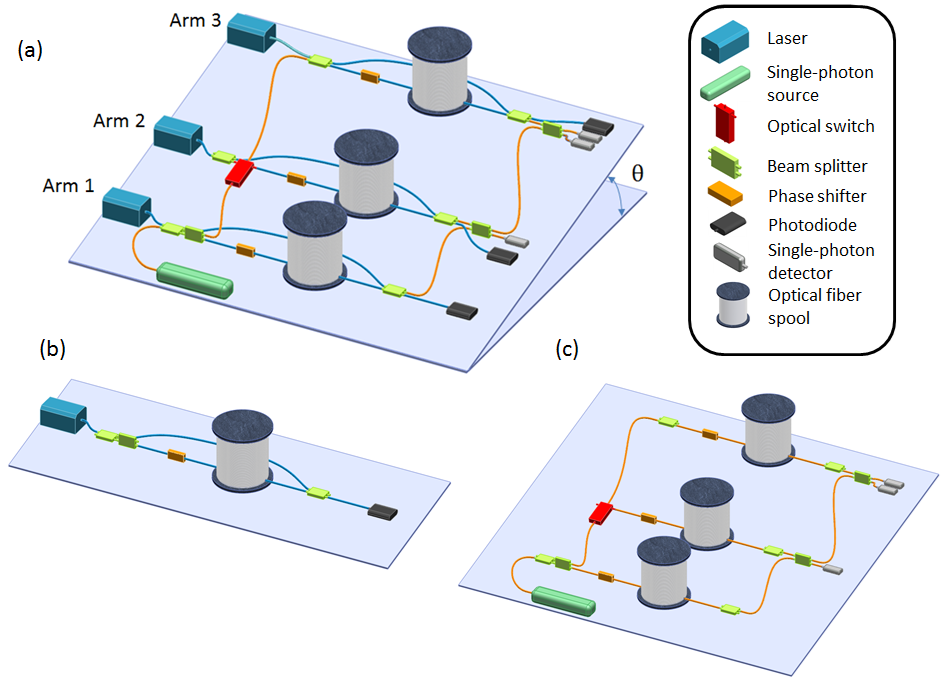}
    \caption{Sketch of the interferometric scheme used to resolve the gravitationally induced phase shift of single-photons. (a) Schematic drawing of the complete setup: the photons (orange and blue paths) are used as interfering particles whereas classical laser light (blue path) is used to stabilize the interferometer by means of an imbalanced MZI in each arm as the angle of inclination ($\theta$) changes. The laser exciting arm 1 can also be used to calibrate the interferometer to operate at the quadrature point in horizontal position ($\theta=0$). (b) Stabilization mechanism: the entire interferometer can be kept at the quadrature point during rotation \textemdash without erasing the gravitational phase information\textemdash by means of an imbalanced MZI in each arm. (c) Phase detection: single photons are coupled into the stabilized interferometer and are used to observe the gravitationally induced phase shift by measuring the difference in photon counts between the detector in arm 2 and the detectors in arm 3 for each angle $\theta$. The high frequency modulation provided by the optical switch creates a time-varying gravitational signal in a low-noise band at the detectors.}
    \label{3armMZI}
\end{figure*}

The required interferometric area for a given wavelength depends on the phase difference as given by Eq. \eqref{PhaseShift}. Due to the small effective mass of the photons, the required area of the  interferometric set-up is much larger than the area needed for matter-wave interferometry \cite{Werner75,Zych12}. For photons in the optical regime ($\lambda_0 \sim10^{-6}$ m), an area of about $10^5$ m$^2$ leads to a phase shift on the order of $\Delta\phi_G\sim 10^{-5}$ rad. In order to achieve such large-area interferometers, optical fibers are ideally suited for compact arrangements even at the table-top scale. Remarkably, commercially available fiber spools containing $100$ km of fiber do not exceed $2*10^{-2}$ m$^3$ of volume and 10 kg of weight which makes them suitable for implementations in the laboratory despite the long path lengths required. \\

\section{MEASURING GRAVITATIONAL PHASE SHIFTS FOR SINGLE PHOTONS}

In a noise-free environment the measurement of the gravitational phase shift would be straightforward. The MZI could first be calibrated by orienting the setup horizontally \textemdash such that the area (\textit{A}) is parallel to the surface of the Earth \textemdash to equalize the optical path lengths of the arms; then, by orienting the interferometer vertically (Fig. \ref{3armMZI}), we could determine the gravitational phase by simply observing the difference in count rates between the detectors. In reality, however, fiber interferometers can also be used as sensors for various physical and chemical variables due to their high sensitivity for external perturbations \cite{Lee03}. This leads in our case to noise induced signal fading \cite{Dandridge82,Cole82}, which is a change in the amplitude of the detected signal as a function of time. It is therefore necessary to design an experiment capable of distinguishing the static gravitational signal from time-dependent noise. Tanaka proposed a possible solution for an all-fiber MZI \cite{Tanaka83}, where the arrangement can be rotated about an axis parallel to its arms. The phase difference between the two arms is angle-dependent and can be measured for all intermediate angles between the horizontal and vertical orientation of the interferometer. Measurements are only taken when the angle between the surface of the Earth and the area of the interferometer is fixed. Due to the static nature of this rotation and the impossibility of calibrating the interferometer for different angles without losing the desired information, this scheme relies heavily on passive stabilization of the fibers. For the proposed area of $5000$ m$^2$ phase-noise larger than $10^{-6}$ rad within the detection band must be suppressed. Although such stability has been  demonstrated in fiber interferometers \cite{Jackson80,Wanser93,Bartolo12}, achieving this precision for such a large-scale interferometer is a challenge. In the scheme we present here, we modulate the gravitational potential difference even when the interferometer is at a fixed position during the measurements ( Fig. \ref{3armMZI}).

\subsection{Setup}
\label{SubSec:Setup}

Our scheme consists of a rotatable 3-arm MZI (Fig. \ref{3armMZI}). Each of the three arms is made up of a fiber beam splitter(FBS), a fiber phase shifter (FPS) and an optical fiber of length $\mathit{l}$. The interferometer can be shielded from external noise, arising from temperature fluctuations and air currents, by placing it in a vacuum chamber. In order to reduce coupling to vibrations, the entire set-up can be placed on an actively stabilized vibration isolation system. An optical switch (OS), consisting of an electro-optical modulator (EOM) and a polarizing beam splitter (PBS), is used to connect the additional arm to the conventional MZI, directing the photons either along arm 2 or arm 3 as a function of time. This technique effectively creates two two-arm MZIs with different spacings between their arms resulting in different gravitational potential differences. Therefore a time-varying signal, modulated at the EOM frequency, is received at the single-photon detectors (SPD) and can be extracted by post-selecting data at the frequency of modulation performed by the EOM. The rotational degree of freedom (e.g. about arm 1, Fig. \ref{3armMZI}) is used to calibrate the interferometer in the horizontal position as well as creating an angle dependent signal by slowly rotating the interferometer and performing measurements at fixed positions.\\

\subsection{Calibration and measurement}
\label{SubSec: CalibrationAndMeasurement}
To avoid signal fading for different angles in the presence of noise without extinguishing the gravitational information in each of the arms, the following strategy may be applied. The interferometer can be calibrated in the horizontal position \textemdash where gravitational differences are absent \textemdash using a frequency stabilized laser source in arm 1 (Fig. \ref{3armMZI}). We can operate at the most phase-sensitive point (quadrature point) of the MZI by adjusting a phase difference of $\frac{\pi}{2}$ between arm 1 and arms 2 and 3. To keep the MZI at this point for all angles and during rotation, the length of each arm is kept constant by means of an additional, imbalanced MZI consisting of the corresponding fiber spool and a short segment of fiber (blue lines in Fig. \ref{3armMZI}) together with a frequency stabilized laser source in each arm. The length of this segment can be chosen to be as short as possible, so that we can assume the noise to be negligible in the passively isolated, noise reducing vacuum chamber. This allows to monitor and thus minimize the noise in the long fiber spool, where the vacuum environment minimizes common-mode noise for each of the three additional interferometers. A linewidth for the stabilizing lasers in the kHz regime is required for good visibility in these unbalanced MZIs. The interferometer can now be slowly rotated (e.g. about arm 1) to change the gravitational potential difference in a controlled way. Because of the active stabilization provided by the imbalanced MZI in each arm, an observer located at one of the rotated arms will measure a stable path length during the entire rotation. After fixing the entire setup at a given angle, single photons \textemdash with frequency different from the lasers \textemdash are coupled into the interferometer via the first FBS (Fig. \ref{3armMZI}c). The single photons will now experience a different gravitational potential depending on which path they take leading to a relative phase shift. Single-photon detectors in arms 2 and 3 can be used to resolve this phase difference. The dynamic EOM-modulation together with the controlled (static) rotation of the interferometer and the active stabilization preserving the calibration condition, allows us to measure the gravitationally induced phase shift on the single photons. In designing this interferometer it is crucial to identify all the physical phenomena that could introduce noise that would swamp the gravitational effect we wish to observe. 

\section{NOISE ANALYSIS}
\label{Sec:NoiseAnalysis}
\subsection{Effects of the rotation of the Earth}
\label{SubSec:EffectsOfEarth}

\begin{figure}
    \includegraphics[scale=0.34]{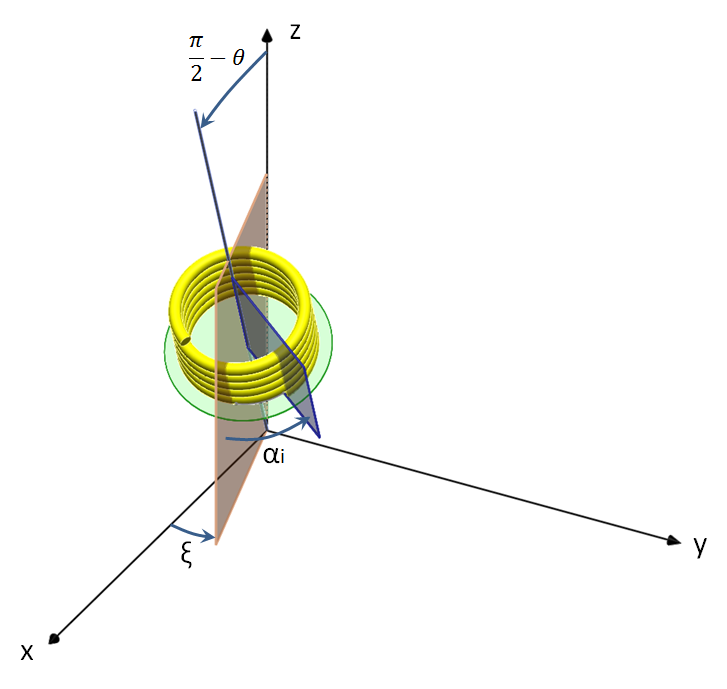}
    \caption{Spool geometry used to calculate the phase-corrections due to the rotation of the Earth. The x-axis is always pointing South, the z-axis is normal to the surface of the Earth and the y-axis is pointing to the East. The photons enter the spool in a plane defined by the angle $\alpha_i$, where \textit{i} is indexing the corresponding fiber spool.}
    \label{Fig:SpoolGeometry}
\end{figure}

The rotation of the Earth introduces an additional relative phase-shift ($\Delta \phi_C$) between the spools. The calibration of the interferometer in the horizontal position effectively sets $\Delta \phi_c$ between the different arms equal to zero. However, as $\theta$ increases, this difference for arms 2 and 3 with respect to arm 1 constantly changes in addition to $\Delta \phi_g$. We can estimate this effect by assuming the velocity of the photon to be constant as seen by an observer in the laboratory. We can then calculate (see Appendix \ref{AppendixA}) the resulting phase-shift between the two outermost arms \textemdash represented by spools 1 and 3 in Fig.\ref{3armMZI} \textemdash within the framework of special relativity to be (up to first order in $\frac{R\Omega}{c}$)

\begin{align}
\label{EQ:DeltaPhiC}
\Delta\phi_c&=\sqrt{1-\frac{b^2 \omega^2+v_z^2}{c^2}}\left(\frac{2 \pi \Delta l}{\lambda}-\frac{2 \pi b R \Omega}{\lambda c}\cos\phi \cdot \right. \nonumber\\
&\left.\Bigg(\cos\xi F(\omega,\alpha_{1,3})+\sin\theta \sin\xi F'(\omega,\alpha_{1,3})\Bigg)\right),
\end{align}
where \textit{b} is the radius of the spool,  $v_z$ is the (average) light speed along the direction of the cylindrical symmetry axis of the spool, $\Delta l\equiv N_1l_1-N_3l_3\equiv L_1-L_3$ is the optical path-length difference  and $\omega$ is the (average) angular speed of the light circling around the spool. The angles $\phi$ and $\theta$ represent the latitude coordinate of the lab on Earth and the spool inclination angle, respectively, as shown in Fig. \ref{Fig:SpoolGeometry}. We define $\xi$ as the angle between the normal projection of the symmetry axis of the spool(s) to the surface of the Earth and the unit vector pointing South. By definition, $\xi=0$ defines a vector pointing South at the location of the lab, whereas $\xi=\pi/2$ defines a vector pointing East. The term linear in $\Delta l$ in Eq. \eqref{EQ:DeltaPhiC} is the usual phase term for standard fiber interferometry, here multiplied by a factor originating from the rotation around the spools axis. Because of the expected value for the gravitational phase ($h=3$ m, $l=10^5$ m) we can calculate a bound on the optical path length difference of about $\Delta l \leq 10^{-11}$m. The oscillating term contains two functions defined by

\begin{align}
\label{EQ:DeltaPhiCSin}
F(\omega,\alpha_{1,3})&:=\sin\left(\frac{\omega L_1}{c}+\alpha_1\right)-\sin\left(\frac{\omega L_3}{c}+\alpha_3\right)\nonumber \\&+\sin(\alpha_3)-\sin(\alpha_1)
\end{align}

\begin{align}
\label{EQ:DeltaPhiCCos}
F'(\omega,\alpha_{1,3})&:=\cos\left(\frac{\omega L_1}{c}+\alpha_1\right)-\cos\left(\frac{\omega L_3}{c}+\alpha_3\right)\nonumber \\
&+\cos(\alpha_3)-\cos(\alpha_1)\,,
\end{align}
that describe an effect arising from the special geometry of the fiber spools. The angles $\alpha_1$ and $\alpha_3$ describe the planes in which the photon enters the fiber spool at time $t=0$ in the laboratory frame. From Eq. \eqref{EQ:DeltaPhiC} \textemdash which can be interpreted as a coupling between the rotation of the Earth and the rotation around the fiber spool \textemdash it is easy to observe, that the amplitude is independent of the length \textit{l} (mod $2\pi b$) of the fiber. Using $b=0.2$ m, $\omega \sim 10^9$ rad s$^{-1}$, $v_z \sim 400$ m s$^{-1}$, $\lambda=1550$ nm and $\phi=48.21^{\circ}$ we can expect a phase-shift on the order of $0.5$ rad. Comparing $\Delta\phi_c$ to the gravitationally induced phase shift (Eq. \eqref{PhaseShift}) by taking the same numbers and the effective refractive index to be $N=1.468$ for a fiber of reasonable length $l=10^5$ m, we expect $\Delta\phi_g$ to be approximately three orders of magnitude smaller. Fortunately, it is relatively easy to align the geometric angles $\alpha_3$ and $\alpha_1$ to be parallel for all spools. For simplicity we consider from now on $\alpha_3=\alpha_1=0$ without loss of generality. We require the effect to be independent of the angle between the fiber spool symmetry axis and the surface normal of the Earth (e.g. between 'horizontal' and 'vertical' position in the laboratory), in contrast to the gravitational effect \footnote{To avoid possible relative pressure induced changes in the refractive index for the fiber spools during rotation, it is beneficial to keep the symmetry axis of the spool and the surface normal to the Earth always parallel}. This can be done by choosing $\xi=n\pi$ ($n\in \mathbb{Z}$), indicating that the axis of rotation between the 'vertical' and the 'horizontal' position should be parallel to the line connecting the cardinal directions West and East. In this case $\Delta\phi_c\propto 2 \sin\Big(\frac{\omega}{2c}\Delta l\Big)\cos\Big(\frac{\omega}{2c}l\Big)$, where we defined $l:=L_3+L_1$. Therefore the oscillating term in Eq. \eqref{EQ:DeltaPhiC} reduces to

\begin{align}
\label{Equ: AngularStability}
\Delta\phi_c^{osc}&=\sqrt{1-\frac{b^2 \omega^2+v_z^2}{c^2}}\frac{4 \pi b R \Omega}{\lambda c}\cos\phi \cdot \nonumber\\
&\sin\Big(\frac{\omega}{2c}\Delta l\Big)\cos\Big(\frac{\omega}{2c}l\Big),
\end{align}
where the argument in the sine function is periodic with a period of $4\pi b$. The periodicity together with the geometry of the model used for calculating Eq. \eqref{EQ:DeltaPhiC} allows for the interpretation of  the sine-term as describing the plane perpendicular to the fiber symmetry axis, where the photon leaves the fiber spool. Note that this term only arises because of the geometry of the fiber spool and would not be observable in a straight fiber interferometer. Setting an upper bound for the maximal angular displacement between the planes of the spools where the photon can leave by using Eqs. \eqref{PhaseShift} and \eqref{Equ: AngularStability}, we obtain a required stability of about $\Delta l \sim 13\,\mu$rad. With a spool radius of about 0.2 m this translates to an angular stability of about 7 mdeg. Though feasible, this is certainly one of the most challenging parts of the experiment. This means that in order to not swamp the gravitational effect by the rotation of the Earth, all of the three possible input planes (described by the $\alpha_i$) and all of the three possible output planes (described by $\Delta l$ and \textit{b}) must not change more than 7 mdeg during rotation from the horizontal \textemdash where this effect is not present due to calibration \textemdash to the vertical.

\subsection{Internal fiber noise}
\label{SubSec:InternalFiberNoise}

\begin{figure}
\centering
    \includegraphics[scale=0.6]{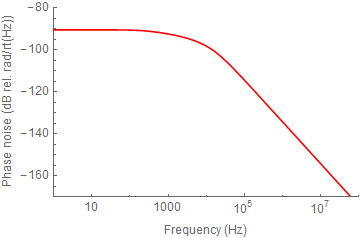}
    \caption{Theoretical curve for the Wanser thermal noise theory. The root-mean-square amplitude of the phase noise fluctuations in the fibers as a function of frequency, as given in \cite{Wanser92}, is shown. The plot represents the expected phase noise for a fiber interferometer operating at a wavelength of $\lambda=1550$ nm and a total length of $2l=200$ km. The used fiber parameters are: thermal conductivity $\kappa=1.37$ W mK$^{-1}$, refractive index temperature coefficient $dn/dT=9.52\times10^{-6}$ K$^{-1}$, effective refractive index $n=1.468$, coefficient of linear expansion $\alpha_L=5\times 10^{-7}$ K$^{-1}$, thermal diffusivity $D=0.82\times10^{-6}$ m$^2$ s$^{-1}$, mode-field radius  $\omega_0=5.2$ $\mu$m and fiber outer radius $a_f=62.5$ $\mu$m. }
    \label{Fig:InternalFiberNoise}
\end{figure}

Another factor that limits the sensitivity and stability of any fiber interferometer is internal noise originating from thermal properties of the fiber itself \cite{Wanser92,Bartolo12,Duan10,Foster07,Dandridge95}. External noise, mainly from temperature fluctuations, air currents and acoustic noise can be greatly reduced by housing the fiber-spool inside an evacuated chamber, shielded additionally from acoustic noise by active vibration isolating systems. In fact, without this shielding it has been shown to be impossible to measure a meaningful signal due to signal fading \cite{Jackson80}. In most applications for short-scale fiber interferometry, the shot noise limit sets a lower bound on the phase-sensitivity of an interferometer. For large-scale fiber interferometers, intrinsic thermal noise of the transmissive medium itself can limit the performance, and must be carefully analysed for the application at hand. There are several theoretical investigations attempting to model the observed noise floors for optical fiber interferometers. The measured power spectral density (PSD) for typical fibers used in optical fiber sensing (see e.g. \cite{Bartolo12,Dong16}), has a $1/f$ dependence in the low frequency regime, which shows good agreement with theories for mechanical dissipation in optical fibers (\cite{Duan10}). For frequencies over 1 kHz, the PSD curve shows excellent agreement with another theory for thermal phase noise by Wanser \cite{Wanser92}. The PSD has a rapid cut-off for Fourier frequencies above 100 kHz, indicating a lower noise contribution in this frequency regime. The root-mean-square amplitude of phase noise fluctuations depends strongly on the geometry and material of the fiber. Using the fiber parameters for Corning's standard single mode fiber SMF-28 \cite{Dong16}, the noise contribution from intrinsic thermal phase noise for an interferometer with total length $2l=200$ km \footnote{because of the optical switch, only two arms of the MZI are involved at a given time instant} can be estimated to be around $10^{-6}$ rad Hz$^{-1/2}$ at 100 kHz according to Wanser's theory (Fig. \ref{Fig:InternalFiberNoise}). A better estimation for this bound can only be given by measuring the crucial parameters for the fiber actually used in this experiment. One can see, that the thermal noise contribution can be tailored by an appropriate choice of signal band-pass filtering at the detectors and proper choice of the modulation frequency with the OS. Passive stabilization by noise damping foundations and vacuum environment are necessary for keeping external noise contributions to a minimum. This might also reduce the need for an active stabilization loop for the short fiber segments connecting the three arms of the interferometer (orange lines in Fig. \ref{3armMZI}). Active stabilization  of the fibers performed by the imbalanced MZIs in each arm keeps the interferometer at the desired quadrature, independent of the inclination angle $\theta$. 

\subsection{Polarization and Dispersion}
\label{SubSec:PolarizationAndDispersion}

Optical fiber interferometers are also sensitive to polarization effects \cite{Kersey88}. The states-of-polarization (SOP) of the two interferometer arm outputs determine the mixing efficiency at the merging beam splitter, where perfect polarization overlap occurs for parallel SOPs resulting in a maximal visibility \cite{Kirkendall04}. The proposed setup relies on rotation around one of the arms of the MZI, so special care must be taken to preserve perfect mixing and avoid polarization drifts and rotations.
One possible approach to overcoming drifts in polarization is to use polarization-maintaining (PM) fibers. The main disadvantages of using PM fibers are the much higher transmission losses ($\sim 0.5$ dB/(km nm) instead of $\sim 0.18$ dB/(km nm)) and the much higher costs. Therefore it is most practical to use standard single-mode fibers with an active stabilization of the polarization in each arm \cite{Kirkendall04}. The required quality of this stabilization depends mainly on the chosen frequency of the optical switch.\\
Because we are aiming for measuring small optical phase shifts with large scale fiber interferometers, dispersive effects might also be of great importance. We assume the single-photon states to have Gaussian spectral amplitude \cite{Loudon00} $f(\omega)=\left(\frac{1}{2\pi\sigma}\right)^{1/4}e^{-i(\omega-\omega_0)t_0-\frac{(\omega-\omega_0)^2}{4\sigma^2}}$, where $\sigma^2$ is the variance of the spectral intensity and $\omega_0$ is the central frequency of the pulse. Allowing for pulses to evolve differently in distinct fibers we can rewrite the detection probability (Eq. \eqref{DetProb}) for a conventional MZI as (Appendix \ref{AppendixB})

\begin{equation}
\label{Equ:DetProbDisp}
P_{\pm} \approx \frac{1}{2} \left( 1 \pm \sqrt{\frac{2\tau\tau'}{\tau^2+\tau'^2}}
e^{-\frac{\Delta \phi_g^2}{4\omega_0^2(\tau^2+\tau'^2)}} \cos(\Delta \phi_g+\phi(t)) \right),
\end{equation}
where $\tau^2\equiv\tau_0^2+\Delta\tau^2$ and $\tau_0$ is the initial temporal width of the photon \footnote{The phase-change due to chirping is too small to be observable and thus neglected.}. The time-domain broadening is given by $\Delta\tau:=D_m l \Delta\lambda$ with $D_m$ representing the dispersion coefficient (ps/(km nm)), \textit{l} the length of the fiber (km) and $\Delta\lambda$ representing the spectral width (nm). The prime in Eq. \eqref{Equ:DetProbDisp} indicates the second fiber with different material coefficient. The Gaussian pre-factor depends on the ratio between path difference, represented by the gravitational phase shift, and pulse length represented by the broadening of the initial pulse in the time domain. It can be shown by using Eq. \eqref{Equ:DetProbDisp} that the dispersion broadening results in an effect that is about two orders of magnitude smaller than the gravitational phase shift for photons with spectral width achievable in recent experiments \cite{Zhong12}. Thus dispersion is not a limiting factor in the proposed experiment.

\subsection{Attenuation and integration time}
\label{AttenuationAndIntegrationTime}

While the interferometer's path lengths should be as large as possible to increase the relative gravitational phase difference between the arms, longer optical fiber paths also introduce more noise and reduce transmission, thus increasing the integration time required for statistically significant measurements. In this setup, the gravitational phase can be resolved by means of a difference in count rates as a function of $\theta$ and the modulation frequency of the optical switch. To determine an upper bound on the required integration time, we assume the single photon source to possess photon statistics following a Poisson distribution at the detectors after passing through the fibers. In order to be visible over the Poissonian noise, the difference in photon counts due to gravity should be at least $\sqrt{\bar{n}_i}$, where $\bar{n}_i$ is the average number of photons registered by the $i^{th}$ detector per time t. Because the optical switch directs the photons either along arm 2 or arm 3, the set-up reduces to a conventional Mach-Zehnder interferometer, where the difference in gravitational potential can be seen as phase change modulated by a phase-shifter. The amplitude in this model is proportional to the potential difference between arms 2 and 3 and the modulation frequency is given by the one of the EOM. The integration time of the detectors can thus be estimated by (see Appendix \ref{AppendixC})

\begin{equation}
\label{MeanPhotons}
t\geq\frac{\bar{N} a \eta P +n_d}{\left(\bar{N} a \eta (A-P)\right)^2}\,,
\end{equation}
where $\bar{N}$ is the number of photons per unit time provided by the single photon source, $\eta$ is the quantum efficiency of the detector, $n_d$ is the dark count rate and $a = 10^{\left(-\frac{\alpha l}{10^4}  - \Sigma_i\alpha_i\right)}$ is the overall attenuation factor of the interferometer with $\alpha$ the attenuation coefficient of the optical fiber, $l$ the length (m) of the horizontal arms of the MZI and the $\alpha_i$ are the various attenuation coefficients of the fiber optic components used in the interferometer. The detection probability for the detector in question at an angle $\theta$ at the quadrature point is denoted by $P$. The detection probability for the \textquoteleft horizontal\textquoteright  position ($\theta=0$), where gravity has no effect, is denoted by $A$. Because of the large area needed to observe even a small gravitational phase shift, it is desirable to work at a wavelength where the optical fibers have high transmission. For standard single-mode fibers this is usually around a wavelength of $1550$ nm. To achieve high quantum efficiencies and low dark count rates for photon detection at this particular wavelength, superconducting nanowire single-photon detectors (SNSPDs) may be used. The dark count rate for SNSPDs have been shown to be as low as 1 Hz for detectors with quantum efficiency $>90\%$ \cite{Marsili13}. Because the integration time is inversely proportional to $\bar{N}$, a single-photon source with high brightness is beneficial. Spontaneous parametric down conversion (SPDC) in non-linear crystals is one of the most versatile and reliable technologies to produce single-photon states of light. The last 20 years have witnessed significant technological improvements in the performance of such sources, which can now reach brightnesses of up to $10^5$ pairs/(s GHz)  per mW of pump with typical bandwidths of the order of 100 GHz \cite{Zhong12}.  By inserting the non-linear crystal in an optical cavity, it is possible to narrow the bandwidth up to 10-100 MHz, while keeping high brightness,  of the order of $10^4$ pairs/(s MHz) per mW of pump power \cite{Luo15}. Typical pump powers that ensure real single-photon regimes range between 1 and 10 mW. Based on these and Eq. \eqref{MeanPhotons}, we expect a maximal required integration time for $\theta=\frac{\pi}{2}$ of about two days.\\

\section{CONCLUSION AND OUTLOOK}

Optical fiber interferometry is a promising technique for table-top experiments aimed at measuring gravitationally induced phase shifts on single photons. Here we have presented a scheme for overcoming the static nature of the gravitational interaction by exploiting a modified Mach-Zehnder interferometer. An additional arm allows us to create a time-varying signal controlled by an optical switch operating at a high frequency chosen to minimize internal fiber noise. By rotating the setup around one of the arms, it is possible to calibrate the interferometer in the horizontal orientation to its most sensitive point of operation. We have shown that by adding an unbalanced MZI in each arm of the MZI, the calibration condition can be preserved during rotation. Single photons injected into the interferometer will therefore be subject to only a gravitational potential difference, depending on the path taken. Due to the fiber spool geometry, the rotation of the Earth can manifest itself as a phase shift large enough to swamp the gravitational phase shift. It is therefore necessary to keep a stringent geometric relation between the various fiber entrance and exit planes during rotation to successfully perform the experiment. Polarization drifts, especially during rotation of the setup, can be avoided by preserving the SOP in each interferometer using an active feedback loop. A successful measurement of the gravitationally induced phase shift would probe the equivalence between the energy of the photon and its effective gravitational mass. While extremely challenging, this experiment has the potential to open the path for table-top experiments capable of testing the interplay between general relativity and quantum mechanics.

\begin{acknowledgments}
We would like to thank A. Sharma, \v{C}. Brukner , M. Zych, R. Weiss, G. Cranch and A. Dandridge for helpful discussions. C.H. is a recipient of a DOC Fellowship of the Austrian academy of Sciences at the Institute of Physics. D.M. is supported by the Kavli Foundation. D.M. and N.M. also gratefully acknowledge support from the United States National Science Foundation. We acknowledge support from the Austrian Science Fund (FWF) through START (Y585-N20), the doctoral program CoQuS, and from the European Commission PICQUE (No. 608062), GRASP(No.613024) and QUCHIP(No.641039) as well as from the University of Vienna via the research platform TURIS.
\end{acknowledgments}

\appendix

\section{Derivation of the phase shift due to the rotation of the Earth}\label{AppendixA}
We now give the derivation for the phase shift arising from the rotation of the Earth (Eq. \eqref{EQ:DeltaPhiC}) to first order in

\begin{equation}
\epsilon\equiv\frac{R\Omega}{c}.
\end{equation}
We consider the photon in a dielectric medium to be a point particle moving with velocity $v<c$. The motion of the photon in a fiber spool is, first, calculated in an earth-centered inertial (ECI) coordinate system. The proper time in this frame is denoted by \textit{t}, and we place the lab initially at the center of the Earth. The Euclidean coordinates of a photon in its spiral motion around the symmetry axis of the spool, located at a distance \textit{h} away from the origin, can be described by the trajectory

\begin{align}
\label{Equ:StartingVector}
\vec{l}(t)&=\begin{pmatrix}
b\,\cos\alpha(t)\\
b\,\sin\alpha(t)\\
v_zt+h\\
\end{pmatrix},
\end{align}
where \textit{b} is the radius of the spool, $\alpha(t)$ is the angle from a predefined plane (Fig. \ref{Fig:SpoolGeometry}) and $v_z$ is the velocity along the symmetry axis of the spool. We now rotate the spool by an angle $\theta'$ about the $y$-axis, as described by the matrix

\begin{align}
R_y(\theta')&=\begin{pmatrix}
\cos \theta' & 0 & \sin \theta' \\
0 & 1 & 0\\
-\sin\theta' & 0 & \cos\theta'
\end{pmatrix},
\end{align}
where $\theta':=\pi/2-\theta$. This is followed by a rotation  about the $z$-axis, which determines the direction of the symmetry axis of the spool, using the rotation matrix

\begin{align}
R_z(\xi)&=\begin{pmatrix}
\cos \xi & -\sin \xi & 0 \\
\sin\xi & \cos\xi & 0\\
0 & 0 & 1
\end{pmatrix}.
\end{align}
We continue by shifting the trajectory to the surface of the Earth with radius \textit{R} by an operator whose action is defined by 

\begin{align}
S(R)&\begin{pmatrix}
a\\
b\\
c
\end{pmatrix}
=\begin{pmatrix}
a\\
b\\
c+R
\end{pmatrix}.
\end{align}
To account for the latitude coordinate we rotate again about the $y$-axis using

\begin{align}
R_y(\phi')&=\begin{pmatrix}
\cos \phi' & 0 & \sin\phi' \\
0 & 1 & 0\\
-\sin\phi' & 0 & \cos\phi'
\end{pmatrix}\,,
\end{align}
where the latitude coordinate is given by $\phi'=\pi/2-\phi$. The rotation of the Earth with an angular speed of $\Omega$ about the $z$-axis of the ECI frame by an angle $\psi(t):=\Omega t+\psi_0$ is described by

\begin{align}
R_z(\psi(t))&=\begin{pmatrix}
\cos \psi(t) & -\sin \psi(t) & 0 \\
\sin\psi(t) & \cos\psi(t) & 0\\
0 & 0 & 1
\end{pmatrix}.
\end{align}
Applying these operators to the vector defined in Eq. \eqref{Equ:StartingVector} results in a world line of the photon given by

\begin{align}
\label{Equ:X}
x(t)&=\Big(ct,R_z(\psi(t))\,R_y(\phi')\,S(R)\,R_z(\xi)\,R_y(\theta')\,\vec{l}(t)\Big) \nonumber\\
&:=\Big(ct,D(t)\vec{l}(t)\Big)
\end{align}
with tangent

\begin{equation}
\label{Equ:XDot}
\frac{d}{dt}x(t)=:\dot{x}(t)=\Big(c,\frac{dD(t)}{dt}\vec{l}(t)+D(t)\frac{d\vec{l}(t)}{dt}\Big)
\end{equation}
and four velocity

\begin{equation}
\label{Equ:FourVelocityECI}
u=\frac{c\dot{x}(t)}{\sqrt{-\eta(\dot{x},\dot{x})}},
\end{equation}
where the Minkowski metric $diag(-1,1,1,1)$ is used. Recall that the photon is considered to be a point particle moving with velocity $v<c$, we can therefore calculate the amount of proper time, say $\tau$ spent in the spool:

\begin{equation}
\label{Equ:ProperTime}
\tau=\int_0^T\frac{d\tau}{dt}dt = \frac{1}{c}\int_0^T\sqrt{-\eta(\dot{x},\dot{x})}dt
\end{equation}
assuming that the photon enters the fiber at $t=0$ and exits at $t=T$. We have also used the relation $-c^2=\eta(u,u)=\left(\frac{dt}{d\tau}\right)^2\eta(\dot{x}(t),\dot{x}(t))=\gamma^2\eta(\dot{x}(t),\dot{x}(t))$, where $\gamma$ denotes the Lorentz factor. In order to calculate \eqref{Equ:ProperTime} we require the photon velocity to be constant as seen by an observer in the lab. Denoting the speed of the photon in the lab by \textit{v} and the four velocity of the lab by $u_L$ we get

\begin{equation}
\eta(u,u_L)=-c^2\gamma(v)=-\frac{c^2}{\sqrt{1-\frac{v^2}{c^2}}}\,,
\end{equation}
or equivalently

\begin{equation}
\frac{v^2}{c^2}=1-\frac{c^4}{\eta^2(u,u_L)}.
\end{equation}
Letting $t_L$ be the lab-proper time, we can use \eqref{Equ:X} together with $b=v_z=h=0$ to find

\begin{equation}
t_L=\Big(1+O(\epsilon^2)\Big)t+C
\end{equation}
for some constant \textit{C}. Hence a constant velocity with respect to the lab frame is equivalent to the requirement that $\eta^2(u,u_L)$ is time-independent to order $\epsilon$:

\begin{equation}
\label{Equ:ConstantVelocityRequirement}
\frac{d}{dt}\eta^2(u,u_L)=0\,.
\end{equation}
The four velocity of the lab is given by 

\begin{equation}
\label{Equ:FourVelocityLab}
u_L=\frac{c\dot{x}_L(t)}{\sqrt{-\eta(\dot{x}_L,\dot{x}_L)}} 
\end{equation}
and can be calculated using \eqref{Equ:X} with $b=v_z=h=0$. Solving Eq. \eqref{Equ:ConstantVelocityRequirement} by using \eqref{Equ:FourVelocityECI} and \eqref{Equ:FourVelocityLab} results in an ordinary differential equation (ODE) with parameters for $\alpha(t)$. The solution takes on the form $\alpha(t)=\alpha(0)+\omega t + \epsilon \alpha_1(t)+\mathcal{O}\Big(\epsilon^2\Big)$ \cite{Teschl12} where $\omega$ is the (average) angular speed of light circling around the spool. After inserting $\alpha(t)$ to order $\epsilon$ in Eq. \eqref{Equ:ConstantVelocityRequirement}) we get

\begin{align}
\label{Equ:AlphaDot}
\dot{\alpha_1}=&-\frac{b^2 \omega^2+v_z^2}{cb}\sin\phi'\Big(\cos\xi\;\cos(\omega t+\alpha_0)+\nonumber\\
&\cos\theta'\,\sin\xi(1-\sin(\omega t+\alpha_0))\Big)\,,
\end{align}
where $\alpha(0)\equiv\alpha_0$ and the approximation $\cos(\alpha(t))=\cos(\omega t+\alpha(0))+O\left(\epsilon\right)$ were used. We can now calculate $\eta(\dot{x},\dot{x})$ by inserting Eq. \eqref{Equ:AlphaDot} and expand the square root in Eq. \eqref{Equ:ProperTime} to first order after factoring out the constant term $\sqrt{1-\frac{b^2 \omega^2+v_z^2}{c^2}}$. The result of this expansion is 

\begin{align}
\tau=&\sqrt{1-\frac{b^2 \omega^2+v_z^2}{c^2}}\Bigg(T-\frac{bR\Omega}{c^2}\sin\phi'\cdot\nonumber\\
&\Big(\cos\xi\sin(\omega T+\alpha_0)+\cos\theta'\,\sin\xi\cos(\omega T+\alpha_0)\Big)\Bigg)\,,
\label{App:Tau}
\end{align}
where $T=\frac{l_iN_i}{c}$ with \textit{i} representing the associated spool with length \textit{l} and group refractive index \textit{N} in the interferometer. The phase difference between arms 1 and 3 of our setup after multiplying by the optical angular frequency is given by 

\begin{align}
\Delta\phi_c&=\frac{2\pi c}{\lambda}\Delta\tau:=\frac{2\pi c}{\lambda}\Big(\tau_1-\tau_3\Big)\nonumber\\
&=\sqrt{1-\frac{b^2 \omega^2+v_z^2}{c^2}}\left(\frac{2 \pi \Delta l}{\lambda}-\frac{2 \pi b R \Omega}{\lambda c}\cos\phi \cdot \right. \nonumber\\
&\left.\Bigg(\cos\xi F(\omega,\alpha_{1,3})+\sin\theta \sin\xi F'(\omega,\alpha_{1,3})\Bigg)\right)\,,
\end{align}
which is equation \eqref{EQ:DeltaPhiC} in Sec. \eqref{Sec:NoiseAnalysis}.
\newline
\\
We finish this appendix with the following calculation, which reduces somewhat the computational complexity of the above. Let us introduce
the following notation

\begin{itemize}
  \item
  $\vec n$ is a Euclidean unit-length vector along the axis of rotation of the Earth;
  \item
  $\vec \ell$ is a Euclidean unit-length vector directed from the center of the Earth to the geometric center of the first coil in the spool;
  \item
  $\vec i$ is a Euclidean unit-length vector along the axis of the spool;
  \item  $\vec j(t)$ is a Euclidean unit-length vector orthogonal to $\vec i$ so that, in Euclidean coordinates in the lab, the position of the photon is
  $$
  t  v_z    \vec i + b \vec j(t)
   \,,
  $$
and where $t\mapsto\vec j(t)$ describes a motion on a flat circle with velocity $\dot \alpha (t)$.
  \item
  $\vec k(t)$ is a Euclidean unit-length vector along $d (\vec j(t))/dt$, thus $\vec i$, $\vec j(t)$ and $\vec k(t)$ are all unit length and pairwise orthogonal, with
  $$
   \frac{d}{dt} \vec j(t) = \dot \alpha (t) \vec k(t)
   \,.
  $$
  \item $U(t)$ denotes a matrix of rotation by angle $\Omega t$ around the axis of the earth. We then have
  $$
    \frac{d}{dt} U(t) = \Omega U(t) \dmat
    \,,
    $$
where $\Omega$ is the rotation velocity and $\dmat =(\dmat_{ij})$ is a time-independent matrix with entries
    $$
     \dmat_{ij}= \epsilon_{ijk} n^k
     \,.
    $$
    It follows that $\dmat$ acts on vectors as a vector product:
    $$
     \dmat (\vec y) \equiv \dmat \vec y  =   \vec y \times \vec n
      \,.
    $$
\end{itemize}
We will only keep track of the leading order corrections in the calculations that follow. Note that there occur some subleading corrections which are $O(\epsilon)$ but \emph{not} of order $\epsilon^2$. In order to isolate the leading order terms, we note that in the new variables the world-line of the photon is
  $$
   x(t) = \big(ct, U(t)(R \vec \ell  + v_z t \vec i + b \vec j(t))\big)
    \,,
  $$
with tangent
\begin{align}
\dot x(t) &= \Big(c, U(t)\big( \Omega \dmat  (R \vec \ell  + v_z t \vec i + b \vec j(t))  + v_z   \vec i +b \dot \alpha \vec k(t)\big)\Big) \nonumber \\
&=\Big(c, c U(t)\big(\epsilon  \dmat  (\vec \ell  +   \frac{v_z t} R \vec i + \frac b R \vec j(t))  + \frac {v_z} c  \vec i + \frac{b \dot \alpha} c \vec k(t)\big)\Big) \nonumber \\ 
&\approx \Big(c, c U(t)\big(\epsilon  \vec \ell  \times \vec n   + \frac {v_z} c  \vec i + \frac{b \dot \alpha} c \vec k(t)\big)\Big)
    \,,
\label{App:RotXDot}
\end{align}
and four velocity $u =c \dot x/ \sqrt{-\eta(\dot x,\dot x)}$. Setting
$$
   \alpha(t) = \omega t + \epsilon \alpha_1(t)
 \,,
$$
and denoting by \textquotedblleft$\cdot$\textquotedblright the Euclidean scalar product and by \textquotedblleft$|~|$\textquotedblright the Euclidean norm, we have
\begin{align}
-\eta(\dot x, \dot x)/c^2 &\approx
1-\big|\epsilon  \vec \ell  \times \vec n   + \frac {v_z} c  \vec i + \frac{b \dot \alpha} c \vec k(t)\big|^2 \nonumber \\
&\approx 1-\frac {v_z^2} {c^2} - \frac{(b \dot \alpha)^2} {c^2}\nonumber \\
&-2 \epsilon  \, ( \vec \ell  \times \vec n) \cdot\big(  \frac {v_z} c  \vec i + \frac{b \dot \alpha} c \vec k(t)\big)\nonumber \\
&\approx 1-\frac {v_z^2} {c^2} - \frac{ (b \omega)^2} {c^2}-2\frac{b^2 \epsilon \omega\dot \alpha_1} {c^2}\nonumber \\
&-2 \epsilon  \, ( \vec \ell  \times \vec n) \cdot\big(\frac {v_z} c  \vec i + \frac{b \omega} c \vec k(t)\big) \; , \nonumber\\
\sqrt{- \eta(\dot x, \dot x)}/c &\approx \sqrt{1
    -    \frac {v_z^2} {c^2} - \frac{(b \omega)^2} {c^2}}
    \Big( 1 \nonumber \\
&-\epsilon \frac{  \frac{b^2 \omega\dot \alpha_1} {c^2} +  ( \vec \ell  \times \vec n) \cdot\big(  \frac {v_z} c  \vec i + \frac{b \omega} c \vec k(t)\big)}{{1-\frac {v_z^2} {c^2} - \frac{(b \omega)^2} {c^2}}}\Big)
 \,.
\label{App:EtaXdotXdot}
\end{align}
Set
\begin{align} 
&\gamma_0^{-1} :=\sqrt{1-\frac {v_z^2} {c^2} - \frac{b^2\omega^2} {c^2}} =: \sqrt{1-\beta_0^2}
 \,, \nonumber \\
&a_1 := \frac {v_z}{c} (\vec \ell  \times \vec n) \cdot \vec i
\,, \nonumber \\
&a_2 \cos(\omega t + \alpha_0) \equiv a_2 \psi(t):=\frac {b\omega}{c} (\vec \ell  \times \vec n) \cdot \vec k (t)
     \,. 
\label{App:Definitions}
\end{align}
Using these variables, we can write
\begin{align}
\sqrt{-\eta (\dot x, \dot x)}/c
  &\approx
   \gamma_0^{-1}\Big( 1 - \epsilon \gamma_0^2 \big( \frac{ b^2 \omega}{c^2} \dot \alpha_1 + a_1 +a_2 \psi\big)\Big)\nonumber
\\
\frac{c}{\sqrt{-\eta (\dot x, \dot x)}}
&\approx \gamma_0 \Big( 1 + \epsilon \gamma_0^2 \big( \frac{ b^2 \omega}{c^2} \dot \alpha_1 + a_1 +a_2 \psi\big)\Big)
   \,.
\label{App:SqrtEtaXdotXdot}
\end{align}
Next, the world-line $x_L$ of the lab is obtained by setting $b=v_z=0$ above:
  $$
   x_L(t) = \big(c t, R U(t)\vec \ell  \big)
    \,,
  $$
with tangent
\begin{align}
  \dot x_L(t) &=
    \Big(c, R\Omega U(t) \dmat   \vec \ell  \Big) =
    \Big(c, c \epsilon U(t)  (\vec \ell  \times \vec n ) \Big)
    \,,
\label{App:LabCoordinates}
\end{align}
and four velocity $u_L =c \dot x_L/ \sqrt{-\eta(\dot x_L,\dot x_L)}$. We have
\begin{equation*}
-\eta(\dot{x}_L, \dot{x}_L) =c^2-c^2 \big|\epsilon  \vec \ell  \times \vec n   \big|^2  \approx c^2
 \,.
\label{App:LabTime}
\end{equation*}
To determine the relative velocity of the photon with respect to the lab we calculate

\begin{align}
-\eta(\dot x, \dot x_L )/c^2 &\approx  1 - \epsilon (\vec \ell  \times \vec n) \cdot \big( \frac{v_z}c \vec i + \frac{b \omega} c \vec k\big)\nonumber\\
& \approx  1 - \epsilon ( a_1  + a_2 \psi)
\,,\\
 -\frac{\eta(\dot x, \dot x_L )}{ \sqrt{\eta(\dot x,\dot x)\eta(\dot x_L,\dot x_L)} }
  & \approx \gamma_0\left(
   1  +  \epsilon \gamma_0^2 \Big(   \frac{b^2 \omega}{ c ^2} \dot \alpha_1 + \right.\nonumber \\
&\left.\beta_0^2 (a_1 + a_2 \psi)\Big)
 \right)
  \,.
\label{App:ConstantVelocityPart2}
\end{align}
A constant velocity of the photon with respect to the lab means that
\begin{equation}
 \label{App:Final}
0 =  \frac{d}{dt}\Big( -\frac{\eta(\dot x, \dot x_L )}{ \sqrt{\eta(\dot x,\dot x)\eta(\dot x_L,\dot x_L)} }
\Big) \approx
 \frac{d}{dt}\Big(
    \frac{b^2 \omega}{ c ^2} \dot \alpha_1 + \beta_0^2  a_2 \psi)\Big)
 \,.
\end{equation}
Integrating this equation once one obtains \eqref{Equ:AlphaDot} at the current order of approximation, leading again to \eqref{App:Tau}.

\section{Derivation of the detection probability for dispersive media}
\label{AppendixB}

\begin{figure}[h!]
\centering
    \includegraphics[scale=0.3]{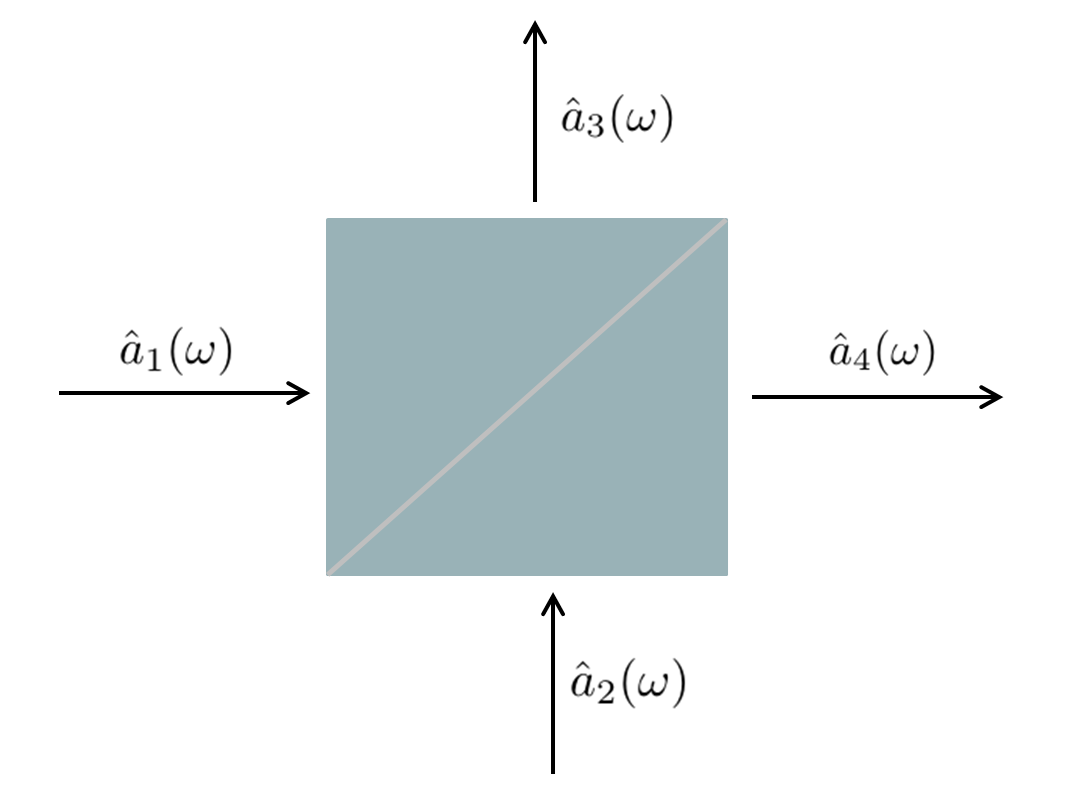}
    \caption{Representation of a symmetric lossless beam splitter showing the annihilation operators associated with the input and output fields.}
    \label{Fig:BeamSplitterRelation}
\end{figure}

The commutation relation for the continuous-mode creation and annihilation operators for modes \textit{i} and \textit{j} is given by \cite{Loudon00}

\begin{equation}
\label{Equ:CommutationRelationOmega}
[\hat{a}_i(\omega),\hat{a}_j(\omega')^{\dagger}]=\delta_{ij}\delta(\omega-\omega')\mathbb{1},
\end{equation}
where $\mathbb{1}$ is the identity operator. A single photon entering the first BS (Fig. \ref{OrdinaryMZI}) is split into a coherent superposition between arm 1 and arms 2 or 3 and recombines at the merging BS. Assigning the annihilation operators for the inputs and outputs of the merging BS as defined in Fig. \ref{Fig:BeamSplitterRelation} we get the relation \footnote{We restrict our analysis to output-mode 4}

\begin{equation}
\label{Equ:BSRelation}
\hat{a}_4(\omega)=T\hat{a}_1(\omega)+R\hat{a}_2(\omega)
\end{equation}
where \textit{R} and \textit{T} are the reflection and transmission coefficients. The spectral width of the wave function is assumed to be small, such that the properties of the BS can be assumed to be independent of frequency. We can calculate the mean photon number in output-mode 4 for an arbitrary input state $\ket{\chi}$ by \cite{Loudon00}

\begin{equation}
\label{App:MeanPhotonNumber}
\braket{\hat{n}_4}=\int dt f_4(t)
\end{equation}
where $f_4(t)\equiv\bra{\chi}\hat{a}_4^{\dagger}(t)\hat{a}_4(t)\ket{\chi}\equiv\braket{\hat{a}_4^{\dagger}(t)\hat{a}_4(t)}$ denotes the mean photon flux. The Fourier-transformed operators $\hat{a}_4(t)$ and $\hat{a}_4^{\dagger}(t)$ are defined by

\begin{align}
\hat{a}(t)&:=\frac{1}{\sqrt{2\pi}}\int_{-\infty}^{\infty}d\omega \hat{a}(\omega)e^{-i\omega t}\label{Equ:A}\\
\hat{a}(t)^{\dagger}&:=\frac{1}{\sqrt{2\pi}}\int_{-\infty}^{\infty}d\omega \hat{a}^{\dagger}(\omega)e^{i\omega t}\label{Equ:B}.
\end{align}

Inserting Eq. \eqref{Equ:BSRelation} into Eq. \eqref{Equ:A} gives

\begin{equation}
\label{Equ:BSRelationTime}
\hat{a}_4(t)=T\hat{a}_1(t)+R\hat{a}_2(t).
\end{equation}

The flux operator is thus given by  

\begin{align}
\hat{a}_4^{\dagger}(t)\hat{a}_4(t)=&\Big(R^{\ast}\hat{a}_2^{\dagger}(t)+T^{\ast}\hat{a}_1^{\dagger}(t)\Big)\Big(R\hat{a}_2(t)+T\hat{a}_1(t)\Big)\nonumber\\
=&\lvert R\rvert^2\hat{a}_2^{\dagger}(t)\hat{a}_2(t)+\lvert T\rvert^2\hat{a}_1^{\dagger}(t)\hat{a}_1(t)\nonumber\\
&+R^{\ast}T\hat{a}_2^{\dagger}(t)\hat{a}_1(t)+T^{\ast}R\hat{a}_1^{\dagger}(t)\hat{a}_2(t),
\end{align}
where an asterix ($\ast$) denotes the complex conjugate of \textit{R} and \textit{T}. The input state is \textemdash due to the action of the first BS \textemdash entangled and given by

\begin{equation}
\ket{\chi}=R\ket{1_f0}+T\ket{01_{f'}}
\end{equation}
which is normalized by the condition $\lvert R\rvert^2+\lvert T\rvert^2=1$. The states $\ket{1_f}$ and $\ket{1_{f'}}$ are defined by

\begin{align}
\ket{1_f}&=\hat{a}_{1f}^{\dagger}\ket{0}=\int dt f(t) \hat{a}_1^{\dagger}(t)\ket{0}\\
\ket{1_{f'}}&=\hat{a}_{1f'}^{\dagger}\ket{0}=\int dt f'(t) \hat{a}_1^{\dagger}(t)\ket{0},
\end{align}
where $\ket{0}$ denotes the vacuum state and $f(t)$ and $f'(t)$ denote the shape of the wave packet in the time domain for the different fibers used as arms in the MZI. In general $f(t)\neq f'(t)$ for different fibers although the input single-photon state into the interferometer is the same. For a single mode we get from Eqs. \eqref{Equ:CommutationRelationOmega}, \eqref{Equ:A} and \eqref{Equ:B} the commutation relation $[\hat{a}(t),\hat{a}(t')^{\dagger}]=\delta(t-t')\mathbb{1}$ that can be used to calculate the mean photon flux for equally transmissive arms

\begin{equation}
\label{Equ:MeanPhotonFlux}
f_4(t)=\lvert R\rvert^2 \lvert T\vert^2\lvert f(t)+f'(t)\rvert^2.
\end{equation}
We assume a Gaussian spectral amplitude for the input single-photon state

\begin{equation}
f(\omega)=\left(\frac{1}{2\pi\sigma}\right)^{1/4}e^{-i(\omega-\omega_0)t_0-\frac{(\omega-\omega_0)^2}{4\sigma^2}}.
\end{equation} 

Fourier transformation gives

\begin{equation}
\label{Equ:SpectralFunction}
f(z=0,t)=\left(\frac{2\sigma^2}{\pi}\right)^{1/4}e^{i\omega_0t}e^{-\sigma^2(t-t_0)^2},
\end{equation}
which describes the shape of the pulse in the time domain at the location of the first beam splitter and where $t_0$ is the time at which the peak of the pulse passes the coordinate origin $z=0$ \cite{Loudon00}. The variance of the intensity given by $\lvert f(z=0,t)\rvert^2$ is $\tau_0\equiv\frac{1}{2\sigma}$. To follow its evolution for $z\neq0$ we can write \cite{Ghatak98}

\begin{equation}
\label{Equ:EvolutionOfPulse}
f(z,t)=\frac{1}{\sqrt{2\pi}}\int_{-\infty}^{\infty} f(\omega) e^{i(\omega t-k(\omega)z+\phi(t))}d\omega\,,
\end{equation}
where $k(\omega)$ denotes the propagation constant and $\phi(t)$ denotes a possible noise term at position \textit{z} and time \textit{t}. Expanding the propagation constant to second order around $\omega_0$

\begin{align}
\label{Equ:PropagationConstant}
k(\omega)=&k(\omega_0)+{\frac{dk(\omega)}{d\omega_0}}(\omega-\omega_0)\nonumber \\
&+\frac{1}{2}\frac{d^2k(\omega)}{d\omega_0^2}(\omega-\omega_0)^2+O\Big(\omega^3\Big)\nonumber\\
=&k_0+\frac{1}{v_g}(\omega-\omega_0)+\frac{1}{2}\rho(\omega-\omega_0)^2+O\Big(\omega^3\Big),
\end{align}
where $\frac{d}{d\omega_0}\equiv\frac{d}{d\omega}\Bigr|_{\substack{\omega=\omega_0}}$, $v_g$ denotes the group velocity of the wave packet and $\rho\equiv\frac{d^2k(\omega)}{d\omega^2}\Bigr|_{\substack{\omega=\omega_0}}$. Inserting this expansion and the function for the spectral amplitude into Eq. \eqref{Equ:EvolutionOfPulse} gives

\begin{equation}
\label{Equ:TemporalFormAtZ}
f(z,t)=\left(\frac{1}{2\pi\tau^2(z)}\right)^{1/4}e^{-\frac{\left(t-\frac{z}{v_g}-t_0\right)^2}{4\tau^2(z)}}e^{i\Phi(z,t)},
\end{equation}
where 

\begin{equation}
\tau^2(z)=\tau_0^2\left(1+\frac{\rho^2z^2}{4\tau_0^2}\right):=\tau_0^2+\Delta\tau^2
\end{equation}
describes the temporal width (measured as square root of the variance) of the photon at a distance \textit{z} away from the origin. The dispersion coefficient as given in the standard form of the temporal broadening for an arbitrary spectral shape  $\Delta \tau=D_m l \Delta \lambda$, is related to the Gaussian shape used in this derivation via $D_m=-\frac{2\pi c\rho}{\lambda^2} 10^6$ (ps/km$\cdot$ nm). The quantity

\begin{align}
\label{Equ:PhaseChirping}
\Phi(z,t)=&\omega_0t-k_0z+\phi(t)-\frac{1}{2}\tan^{-1}\left(\frac{\rho z}{2\tau_0^2}\right)\nonumber\\
&+\frac{\rho z \left(t-\frac{z}{v_g}-t_0\right)}{8\tau_0^2\tau^2(z)} \nonumber \\
&\equiv\omega_0 t-k_0z+\phi(t)+\Xi\,,
\end{align}
represents the phase, where the term inversely proportional to $\tau(z)^2$ leads to the phenomenon of chirping \cite{Ghatak98}.
Using Eq. \eqref{App:MeanPhotonNumber}, inserting Eq. \eqref{Equ:TemporalFormAtZ} into Eq. \eqref{Equ:MeanPhotonFlux} and neglecting $\Xi$ in the phase term due to its negligible contribution as compared to the gravitational phase results for different broadening in the two noise-reduced fibers in Eq. \eqref{Equ:DetProbDisp} of Sec. \eqref{Sec:NoiseAnalysis}

\begin{equation}
P_{\pm} \approx \frac{1}{2} \left( 1 \pm \sqrt{\frac{2\tau\tau'}{\tau^2+\tau'^2}} e^{-\frac{\Delta \phi_g^2}{4\omega_0^2(\tau^2+\tau'^2)}} \cos(\Delta \phi_g+\phi(t)) \right),
\end{equation}
where $\Delta\phi_g\equiv\frac{\Delta l}{v_g}$ in this context.

\section{Derivation for the estimated integration time}
\label{AppendixC}

The single-photon source is assumed to produce on average $\bar{N}$ photons per unit time interval. We consider symmetrical beam splitters where a single photon can be found in both output modes with equal probability. The arms of the MZI are assumed to be equally transmissive and the attenuation coefficient can be calculated from $\Sigma_i\alpha_i+\alpha\,l=10\,\text{log}_{10}\frac{P_{in}}{P_{out}}$ with \textit{l} in \textit{km} and $P_{in}$ and $P_{out}$ representing the input and output power, respectively. After the merging BS there are now $\bar{N}\cdot a$ photons detectable per unit time. The detector has a probability of $P$ to observe a photon with a quantum efficiency of $\eta$ and a dark count rate of $n_d$. Thus the total number of photons expected to produce a signal at the detector after a time \textit{t} is given by

\begin{equation}
\label{Equ:PhotonsAtDetector}
\bar{n}(t)\equiv\left(\bar{N}\,a\,\eta\,P+n_d\right)t.
\end{equation}
To estimate the integration time, we assume that we are Poisson noise limited, resulting in the inequality

\begin{equation}
\label{Equ:PoissonNoiseInequality}
\sqrt{\bar{n}(t)}\leq\bar{n}_{sig}(t).
\end{equation}
If we denote the detection probability in the absence of gravitational effects by \textit{A}, we can calculate the expected number of photons indicating the gravitational phase shift within a time interval \textit{t} by $\bar{n}_{sig}(t)=\left(\bar{N}\,a\,\eta\,(A-P)\right)t$. Solving Eq. \eqref{Equ:PoissonNoiseInequality} for \textit{t} results in an estimated time after which the gravitational effect is visible over Poissonian noise given by

\begin{equation}
t\geq\frac{\bar{N} a \eta P +n_d}{\left(\bar{N} a \eta (A-P)\right)^2}.
\end{equation}
In order to give a quantitative number for \textit{t} we denote the single-photon detector at the end of arm 2 by D1 and the detectors at the end of arm 3 by D2 and D3 respectively. Because of the optical switch \textemdash effectively creating two 2-arm MZIs \textemdash and the interferometer being calibrated to the quadrature point in the horizontal orientation, the probability of detecting a photon in either detector depends on the arms composing the MZI at a given time. In particular, if the noise is suppressed with the methods described in the main text, the detection probability  for D1 is $1/2$  (arms 1 and 2) or $1/4$  (arms 1 and 3). Similarly the probability of detecting a photon in D2 or D3 is $1/4$ if arm 2 is open and $3/8$ if arm 2 is closed. The integration time is therefore estimated by using A=$1/4$ in order to get a lower bound. To calculate P we can use Eq. \eqref{DetProb} with $1/2$ replaced by $1/4$ for the 3-arm MZI, $\Delta\phi_g$ as given by Eq. \eqref{PhaseShift}, $\phi(t)=\pi/2$ to account for the noise suppressed quadrature point, $l=10^5$ m, $h=1$ m $N=1.468$ and $\lambda=1550$nm. Using $\bar{N}=10^6$ s$^{-1}$, $\alpha=17$ dB km$^{-1}$, $\Sigma_i\alpha_i=0.5$ dB, $\eta=0.9$ and $n_d=1$ s$^{-1}$ we can calculate the estimated time (Eq. \eqref{MeanPhotons}) for the gravitational effect to be visible over Poissonian noise to be almost 2 days. For other inclination angles the integration time is substantially longer, e.g. for $\theta=\pi/2$ and equal values for the other parameters the integration time is almost 4 days.

\include{FinalVersion.bib}
\bibliographystyle{Science}

\end{document}